\documentclass[11pt,twoside]{article} 
\usepackage{asp2004}
\usepackage{epsf}
\usepackage{psfig}
\usepackage{lscape} % for landscape figures/tables

\begin{document} 
\title{The mass of the sdB primary of the binary HS~2333+3927}
\author{U. Heber, H. Drechsel, C. Karl} 
\affil{Dr. Remeis-Sternwarte, Astronomisches Institut
der Universit\"at Erlangen-N\"urnberg, Sternwartstra\ss e 7,
D-96049 Bamberg, Germany}
\author{R. {\O}stensen, S. Folkes} 
\affil{Isaac Newton Group of Telescopes, E-37800 Santa Cruz de La Palma,
Canary Islands, Spain}
\author{R. Napiwotzki} 
\affil{Department of Physics \& Astronomy, University of Leicester, 
University Road, Leicester LE1 7RH, UK }
\author{M. Altmann} 
\affil{Departamento de Astronomia, Universidad de Chile, Camino El
Observatorio 1515, Las Condes, Chile}
\author{O. Cordes} 
\affil{Sternwarte der Universit\"at, Auf dem H\"ugel 71, D53121 Bonn, Germany}
\author{J.-E. Solheim} 
\affil{Institute of Theoretical Astrophysics
University of Oslo, p.box 1029, N-0315 Blindern-Oslo, Norway}
\author{B. Voss, D. Koester} 
\affil{Institut f\"ur Theoretische Physik und Astrophysik, Universit\"at 
Kiel, 24098 Kiel, Germany}
% please use mixed case type and Initials-Familyname order
% The last author's name must be proceeded by 'and'.
% If you have many authors and institutions you may want to use
% superscripts instead and collect all institutions at the end of the
% author list. In this case the superscript should follow the
% separating comma 
% \author{A. Author1,$^1$ B. Author2,$^2$ C. Author3,$^3$ 
% and D. Author4$^3$}
% \affil{$^1$Institute1 with Postal Address\\
% $^2$Institute2 with Postal Address\\
% $^3$Institute3 with Postal Address\\}
\begin{abstract} 
Short period sdB binaries
with cool companions
are crucial to understand pre-CV evolution, because they will evolve into
cataclysmic variables, when the sdB will have left the extended horizontal branch. 
Recently we discovered the sixth such system, HS~2333+3927, 
consisting of an sdB star and an M dwarf (period: 0.172~d) with 
a very strong reflection effect, but no eclipses. 
%Therefore additional information, such as a spectroscopic gravity
%measurement are required to derive an unambiguous solution.
The reflection is stronger than in any of the other similar systems which 
renders
a quantitative spectral analysis very difficult because the Balmer line
profiles may be disturbed by the reflected light.
A spectroscopic analysis 
results in $ {T_{\rm eff}}$ = 36\,500\,K, $\log{g}$ = 5.70, and 
$\log (n_{\rm He}/n_{\rm H}) = -2.15$. 
%These characteristics are typical for 
%sdB stars. 
Mass-radius relations were derived from the results of the analysis
of light and radial-velocity curves. Comparison with the mass-radius
relation derived from the surface gravity of the sdB star favours a rather 
low mass of 0.38~\mbox{$\rm M_{\odot}$}\ for the primary. 
The mass of the companion is
0.29~\mbox{$\rm M_{\odot}$}. 
%A mass of 0.29\Msolar\ for the dM companion is
%derived using an observed  
%mass-radius relation of lower main sequence stars.
HS~2333+3927 is the only known sdB+dM
system with a period above the CV period gap. 
\end{abstract}

\section{Introduction}
% use \boldmath if you include mathematical expressions
% Capitalize significant words in the section (and subsection,
% subsubsection) titles
% Do not use \\ to start a new paragraph, but rather \par. 
SdB binaries are important to clarify the evolutionary origin of sdB stars
because
the analysis of light and radial velocity curves can constrain their
dimensions and masses. However, only five suitable systems are known up to
now.

%Subluminous B (sdB) stars dominate the populations of faint blue stars of
%our own Galaxy and are found in both the old disk and in halo populations,
%e.g. as stars forming the blue tails to the horizontal branches of globular
%clusters. However, important questions remain over the evolutionary paths
%and the 
%appropriate time scales. 
There is general consensus that the sdB stars can
be identified with models for Extreme Horizontal Branch (EHB) stars
(Heber 1986). Like all HB stars they
are core helium burning objects. However, their internal structure differs
from typical HB stars, because their hydrogen envelope is very thin
($<$1\% by mass) and therefore inert. As a consequence EHB stars evolve
directly to the white dwarf cooling sequence, thus avoiding a second red
giant
phase.
How they evolve to the EHB configuration is controversial.
The problem is how the mass loss mechanism of the progenitor manages
to remove all but a tiny fraction of the hydrogen envelope at {\em
precisely} the same time as the He core has attained the mass 
($\sim$0.5~\mbox{$\rm M_{\odot}$}) required for the He flash.

Considerable evidence is accumulating that a significant fraction of the 
sdB stars reside in close
binaries (Maxted et al.~2001; Napiwotzki et al., 2004, see also Karl et al,
these proceedings).
Therefore mass transfer should have played an important role in the evolution
of such binary systems. Detailed investigations of sdB binaries, in
particular eclipsing systems, are crucial to determine their masses.
However, only three such eclipsing binaries, HW~Vir, PG~1336-013, and
HS~0705+6700 (see Drechsel et al. 2001) and two non-eclipsing ones
are known up to now, which consist of an sdB star
and an M dwarf companion revealed by reprocessed light from the primary
(reflection effect). Recently, Heber et al. (2004)
discovered
another related system, HS~2333+3927, which is, however,
not eclipsing, but otherwise possesses very similar system parameters
and configuration.

\section{Observations}

The first hint for light variations emerged when HS~2333+3927 was monitored
at the Nordic Optical Telescope on October 19, 1999 in order to search for
pulsations.
The light curve of HS~2333+3927 (see Fig. 1) was measured from CCD 
photometry in 
the $B$, $V$, and $R$ bands at four telescopes (JKT 1.0m, Calar Alto 1.23~m, 
IAC 0.8~m \& Hoher List 1.06~m) in 15 nights between July 2 and November 11, 
2002. 

The radial velocity curve (see Fig. 2) was measured 
during an observing run at the Calar Alto Observatory with the TWIN 
spectrograph at the 3.5m telescope. A total of 18 
spectra were taken from 11 to 18 August 2002 covering the 
wavelength ranges from 3900~\AA {} to 5000~\AA {} at a
resolution of 1.3~\AA {} in the blue part of the spectrum, and from
6000~\AA {} to 7000~\AA {} at a resolution of 1.2~\AA {} in the red part.

HS~2333+3927 was also observed at the
Calar Alto Observatory 
with the CAFOS spectrograph at the 2.2m telescope in order to derive 
atmospheric parameters by Balmer line fitting. A total of 13 low 
resolution spectra were taken on August 31, 2003 covering 
the wavelength range from 3300~\AA {} to 6000~\AA {} 
at a spectral  resolution of 4.5~\AA. This allowed to measure
the entire Balmer series (except H$\alpha$) up to the Balmer jump.

\section{Analysis of light and radial velocity curves}

The numerical solution of the light curves was performed with the
Wilson-Devinney (1971) based light curve program MORO (Drechsel et al. 1995).
The best fit to the $B$, $V$, $R$ light curves is shown in Fig.1.
A strong reflection effect is visible ($\Delta B$~=~0.21, $\Delta V$~=~0.28,
$\Delta R$~=~0.33~mag), but no eclipses are apparent. 
We determined the period to be $P$=0.1718023~d.

The spectrum of HS~2333+3927 is single-lined and the radial velocity curve
is sinusoidal (Fig.~2) indicating that the orbits are circular. 
The semi-amplitude is $K_1 = 89.6$ km/s and 
the mass function follows as $f(m)$~=~0.0128\,M$_{\odot}$.

The light curve analysis provided us with an estimate of the system's
inclination. For any assumed primary mass we can calculate the secondary
mass from the mass function. In addition the orbital radius $a_1$
of the primary can be
calculated.
The separation $a$ of the components can then be calculated
from the mass ratio q.
The light curve allows us to determine the ratio of the radii in units of
the
separation. Hence we can determine the radii of both components for any
given sdB mass, i.e we can derive mass-radius relations for both components.

\begin{figure}
\vspace{8.0cm}
\includegraphics{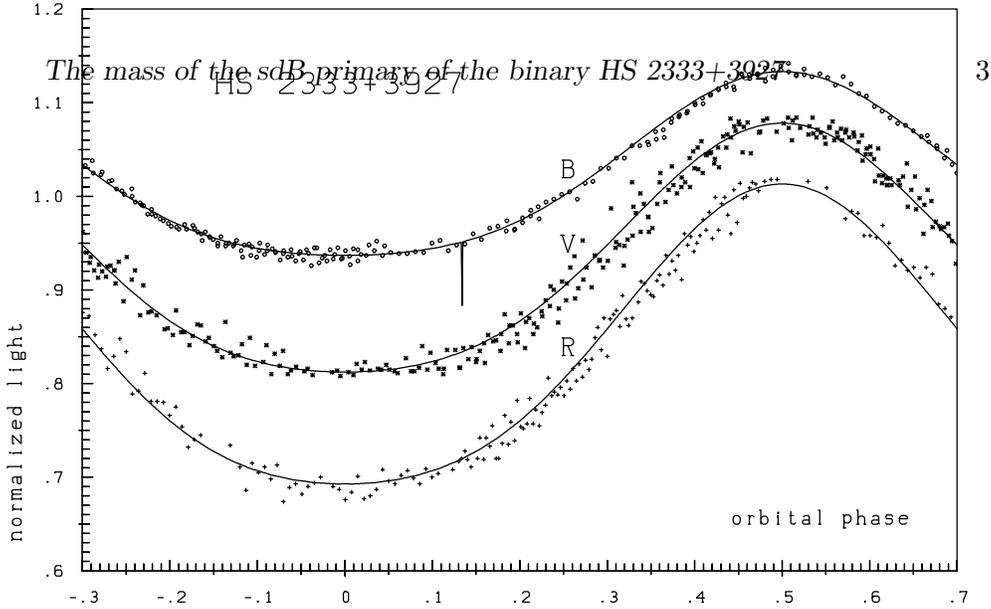}
\caption[]{$BVR$ light curves of HS~2333+3927 (Heber et al. 2004).}
\end{figure}

\begin{figure}
\vspace{7.4cm}
\includegraphics{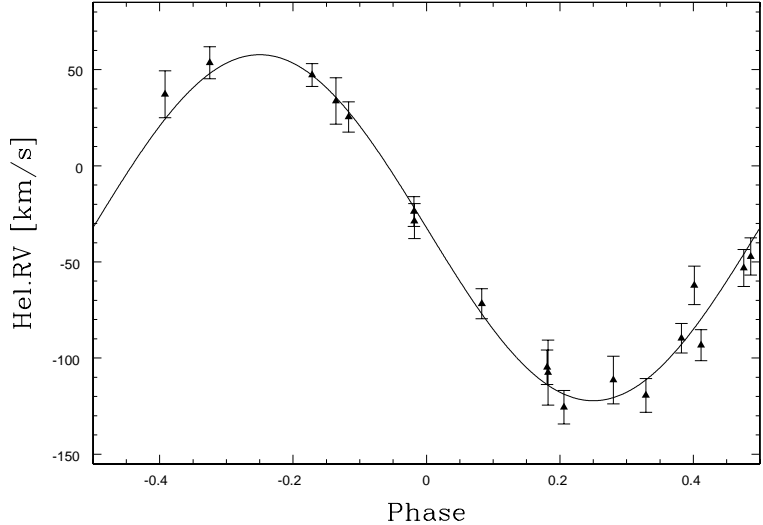}
\caption[] {Radial velocity curve of HS~2333+3927 (Heber et al. 2004).}
\end{figure}

\section{Spectroscopic Analysis}

The Balmer and helium lines in the blue spectra can be used to determine the 
atmospheric parameters by performing a quantitative spectral analysis. 
Because the medium resolution TWIN spectra cover only few of the Balmer
lines, 
we used the CAFOS spectra which cover the Balmer series to its limit. Since 
the high Balmer lines are probably least affected by reprocessed light from 
the secondary, much
emphasis was put in the fit procedure to reproduce these lines well. 
Since the spectrum displays helium lines from two stages of ionization, we
made also use of the ionization equilibrium of helium to determine 
${T_{\rm eff}}$.

The atmospheric parameters determined from spectra taken 
near lower conjunction should be least affected by reprocessed light from 
the secondary and, 
therefore, the atmospheric parameters derived from them should be the closest
approximation to the true atmospheric parameters of the sdB star.

Based on the above considerations we derived
$ {T_{\rm eff}}$\ = 36\,500 $\pm$ 1000~K, $\log{g}$ = 5.70 $\pm$ 0.1
and $\log (n_{\rm He}/n_{\rm H}) = -2.15 \pm 0.15$ for the atmospheric
parameters of HS~2333+3927. 

\section{Masses and Radii}

The mass-radius relations derived from the analysis of the light and radial
velocity curves can be compared to independently determined relations. 
For the sdB star such a relation follows from the gravity (Newton's law). 
For the cool companion a theoretical mass-radius relation for M-dwarfs may be
appropriate. In principle the masses and radii of the sdB star and the M
dwarf, respectively, could be determined
by requesting these mass-radius relations to be consistent.

Fig.~3a compares the mass-radius relation of the sdB derived
from the light and radial velocity curves
to that derived from gravity. The gravity of
$\log{g}$~=~5.7 is too low to match the
M-R-relation from light and radial velocity curve for reasonable masses.
The intersection of both M-R-relations in Fig. 3a would give a sdB mass of
0.2~\mbox{\,$\rm M_{\odot}$}, clearly too low for a core helium burning star.
Evolutionary scenarios by Han et al. (2003) suggest that
the most likely mass is $\approx$0.47~\mbox{$\rm M_{\odot}$}, 
but possible masses range
from 0.38 to 0.8~\mbox{\,$\rm M_{\odot}$}.
As can be seen from Fig.~3a the sdB gravity should be
$\log{g}$~$\approx$~5.86 if the sdB star is of canonical mass.
This is significantly higher than derived from a quantitative spectral analysis
of the
Balmer lines: $\log{g}$~=~5.7$\pm$0.1. However, if we adopt the lowest mass 
core helium burning star that can form in the Han et al. scenario, consistency 
could be achieved if we adopt the largest gravity allowed by the spectroscopic 
analysis (see Fig.~3a).

The mass-radius relation for the companion star is compared to observations
and model
predictions for M-type main sequence stars in Fig.~3b.
The empirical relation for the unseen companion of HS~2333+3927 lies
above those for normal M-stars. However, it is
possible that the M-star is over-luminous due to the strong irradiation of its
surface by the nearby hot star and hence has a larger radius than a normal
M dwarf of the same mass.

%-- mr_sdb.EPS -----------------------------------------------------------
\begin{figure}
\vspace{7.5cm}
\includegraphics{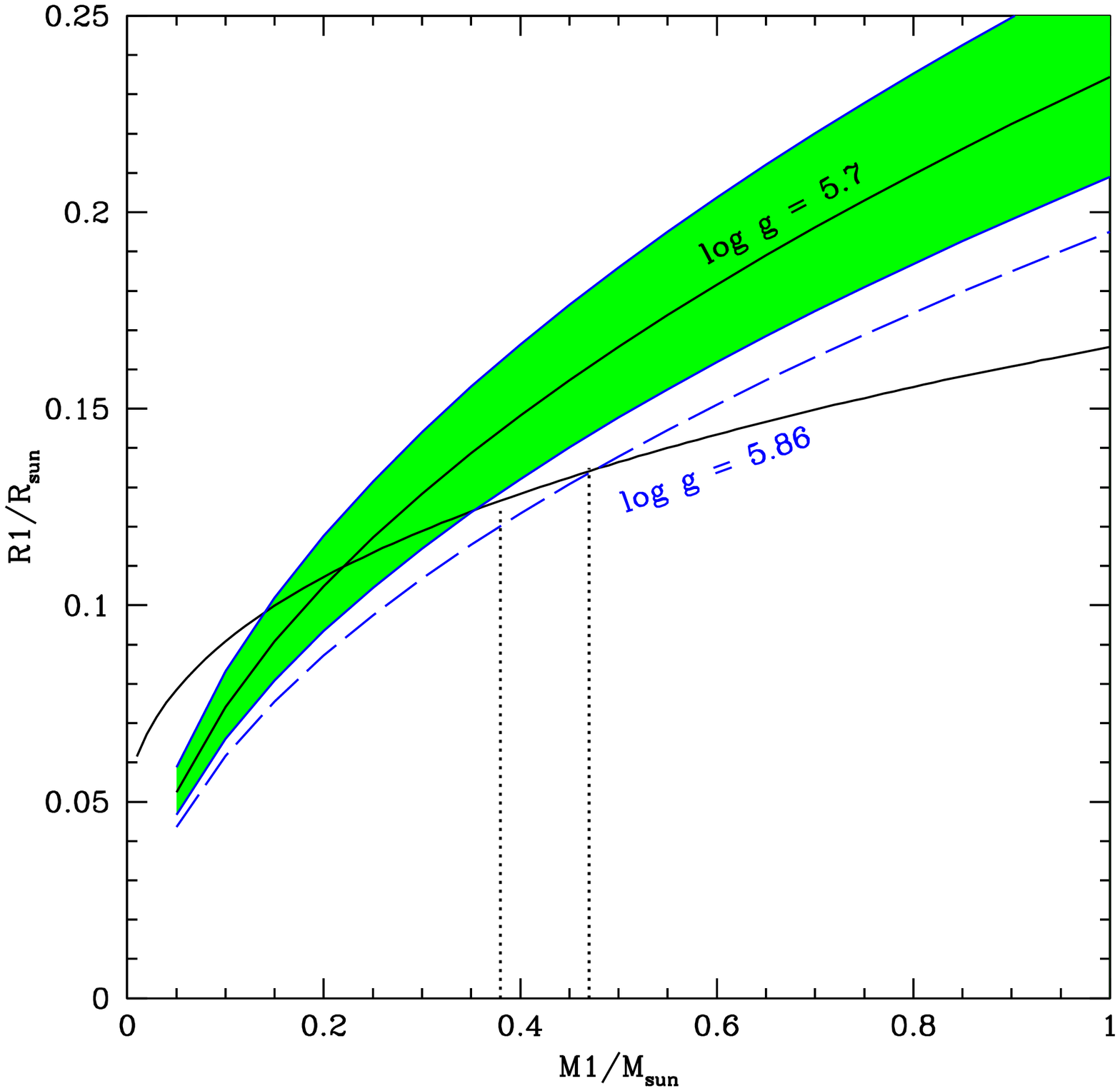}
\includegraphics{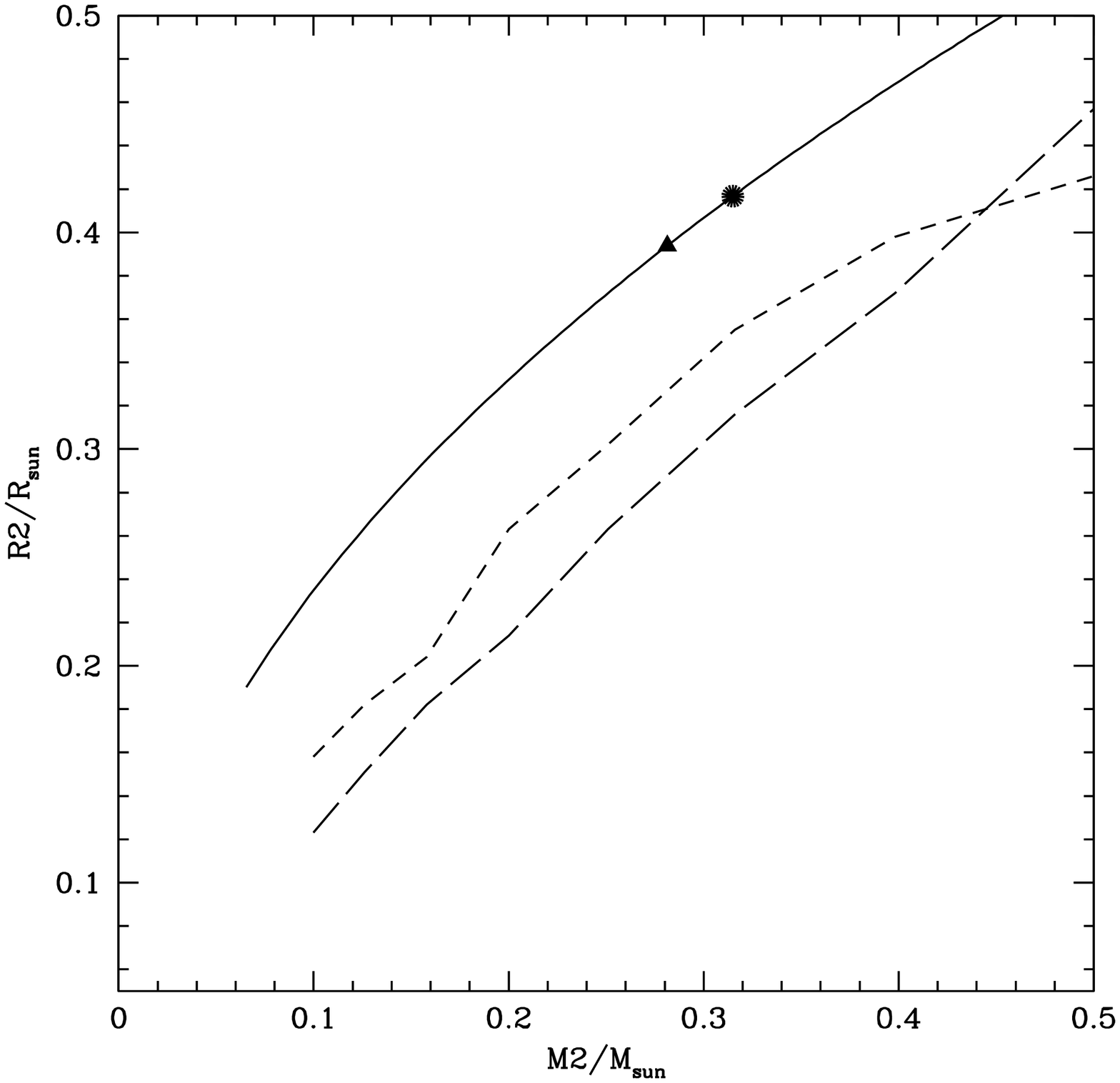}
\caption[]{a) left: Comparison of the sdB mass-radius relation
from the analysis of light
and radial velocity curves to those derived from different gravities. The
spectroscopic $\log{g}$ estimate is 5.7$\pm$0.1. Dashed lines mark the most
probable (0.47~\mbox{$\rm M_{\odot}$})
and the lowest mass (0.38~\mbox{$\rm M_{\odot}$} according to the 
evolutionary models of
Han et al. (2003). Note that the gravity would need to be as high as 5.86 
if the star were of canonical mass (0.47~\mbox{$\rm M_{\odot}$}).\\
b) right: Comparison of the companions mass-radius relation derived from
the analysis of light
and radial velocity curves to  relations for M-type
dwarfs (short-dashed: observed relation from Clemens et al. 1998; 
dashed: theoretical predictions from  Baraffe \& Chabrier,
1996). Filled circle: sdB has canonical mass (0.47~\mbox{$\rm M_{\odot}$}); 
triangle: sdB has lowest possible mass.
%from {\it Allen's Astrophysical quantities, 4$^{th}$ed.}).
}
\label{mr_sdb}
\end{figure}

\section{Discussion}

\subsection{Results}

From the analysis presented above we conclude that a model for a low mass
($\approx$0.38~\mbox{$\rm M_{\odot}$}) core helium burning sdB star and a 
0.29~\mbox{$\rm M_{\odot}$}\ M-dwarf
fits the observations best.
However, the results have to be taken with a grain of salt.
The spectral analysis of the Balmer lines,
is plagued by reflected light, as is evident from the
line profile variations observed in H$\alpha$, but less
obvious in other Balmer lines. An improved measurement of the gravity,
therefore, is urgently needed.

Another important observational constraint would be a precise measurement of
the projected rotational velocity.
The rotation of the sdB star in HS~2333+3927  is very likely
tidally locked to the orbital motion. Drechsel et al. (2001) showed this to
be true for HS~0705+6700.
Determining the projected rotation velocity would
therefore allow an independent estimate of the inclination of HS~2333+3927.

The spectral lines of the Lyman series are sensitive gravity indicators and
plenty of metal lines can be used to measure $v\,\sin{i}$. 
These measurements will
allow us to constrain mass and radius much better

\subsection{SdB stars and pre-CV Evolution}

Short period sdB binaries
with main sequence companions, like HS~2333+3927, are important not only to
understand the formation and evolution of sdB stars. When the sdB star will
have left the EHB, it will evolve into a cataclysmic variable.
Therefore, these objects
are also crucial to understand pre-CV evolution. 
Schenker (these proceedings) has suggested that cataclysmic variables with
short periods, i.e. below the CV period gap evolve from sdB + dM binaries.
While four other known
systems have periods between 1.7~h to 2.8~h, i.e. below the CV period 
gap, the period (4.13~h) of HS~2333+3927 is a bit above the CV period 
gap. Its orbital separation is so small that the secondary is appreciably 
distorted. The shrinkage of the binary orbit 
by gravitational wave radiation will initiate mass transfer turning the
system into a cataclysmic variable. It is likely that the HS~2333+3927 system
will then still have a period larger than the CV period gap. 
Hence sdB + dM binaries form an evolutionary channel to 
the population of longer period cataclysmic variables as well as to the 
short-period population.

%\subsection{The First Subsection: Example for a Simple Table}
%
%\begin{table}[!ht]
%\caption{A simple table}
%\smallskip
%\begin{center}
%{\small
%\begin{tabular}{ccccc}
%\tableline
%\noalign{\smallskip}
%Component & Velocity & $N_{\mathrm{O\, VI}}$ & $N_{\mathrm{H}}$ & Covering factor\\
%          &[km s$^{-1}$]& [cm$^{-2}$]& [cm$^{-2}$]& \\
%\noalign{\smallskip}
%\tableline
%\noalign{\smallskip}
%1 & $-1352$ & $1.7 \times 10^{15}$ & $9.0 \times 10^{14}$ & 1.0\\
%2 & $-599$  & $4.1 \times 10^{15}$ & $2.1 \times 10^{15}$ & 0.9\\
%3 & $-792$  & $2.1 \times 10^{15}$ & $1.7 \times 10^{15}$ & 0.7\\
%4 & $-1029$ & $6.0 \times 10^{15}$ & $2.3 \times 10^{15}$ & 0.5\\
%\noalign{\smallskip}
%\tableline
%\end{tabular}
%}
%\end{center}
%\end{table}

%\subsubsection{The first subsubsection: Figures}

\acknowledgements{C.K. gratefully acknowledges financial support by the
conference organizers. M.A. is financially supported by FONDAP~1501~0003.}

\end{document}